*Article*

# Coupling between Spin and Charge Order Driven by Magnetic Field in Triangular Ising System LuFe₂O₄₊δ

**Lei Ding** [1,*], **Fabio Orlandi** [1], **Dmitry D. Khalyavin** [1], **Andrew T. Boothroyd** [2], **Dharmalingam Prabhakaran** [2], **Geetha Balakrishnan** [3] and **Pascal Manuel** [1,*]

[1] ISIS Pulsed Neutron Facility, Rutherford Appleton Laboratory, Harwell Oxford, Didcot OX11 0QX, UK; fabio.orlandi@stfc.ac.uk (F.O.); dmitry.khalyavin@stfc.ac.uk (D.D.K.)

[2] Department of Physics, Oxford University, Clarendon Laboratory, Parks Road, Oxford OX1 3PU, UK; Andrew.Boothroyd@physics.ox.ac.uk (A.T.B.); dharmalingam.prabhakaran@physics.ox.ac.uk (D.P.)

[3] Department of Physics, University of Warwick, Coventry CV4 7AL, UK; G.Balakrishnan@warwick.ac.uk

\* Correspondence: lei.ding@stfc.ac.uk (L.D.); pascal.manuel@stfc.ac.uk (P.M.); Tel.: +44-123-544-6290 (P. M.)

**Abstract:** We present a study of the magnetic-field effect on spin correlations in the charge ordered triangular Ising system LuFe₂O₄₊δ through single crystal neutron diffraction. In the absence of a magnetic field, the strong diffuse neutron scattering observed below the Neel temperature ($T_N$ = 240 K) indicates that LuFe₂O₄₊δ shows short-range, two-dimensional (2D) correlations in the FeO₅ triangular layers, characterized by the development of a magnetic scattering rod along the 1/3 1/3 L direction, persisting down to 5 K. We also found that on top of the 2D correlations, a long range ferromagnetic component associated with the propagation vector $\mathbf{k_1}$ = 0 sets in at around 240 K. On the other hand, an external magnetic field applied along the c-axis effectively favours a three-dimensional (3D) spin correlation between the FeO₅ bilayers evidenced by the increase of the intensity of satellite reflections with propagation vector $\mathbf{k_2}$ = (1/3, 1/3, 3/2). This magnetic modulation is identical to the charge ordered superstructure, highlighting the field-promoted coupling between the spin and charge degrees of freedom. Formation of the 3D spin correlations suppresses both the rod-type diffuse scattering and the $\mathbf{k_1}$ component. Simple symmetry-based arguments provide a natural explanation of the observed phenomenon and put forward a possible charge redistribution in the applied magnetic field.

## 1. Introduction

The layered oxide LuFe₂O₄, bearing significant spin and charge frustration in a triangular lattice, has attracted wide interest thanks to reported high-temperature multiferroic properties, pivotal for technological applications in spintronics [1–6]. Such properties have been related to the charge ordering of Fe²⁺ and Fe³⁺ cations, which appears as a three-dimensional (3D) long range ordering below 320 K [2]. An apparent coupling between polarization and magnetism has also been previously proposed based on the observation of significant changes in the electric polarization when spin ordering sets in at $T_N$ = 240 K [3]. Although the polar nature of the charge ordering was not confirmed by the later structural study [7], LuFe₂O₄ remains a focus of intensive investigations as a model frustrated system.

The parent structure of rhombohedral LuFe₂O₄ (space group *R-3m*) (Figure 1) is characterized by the stacking of three FeO₅ face-sharing bipyramidal layers along the c-axis separated by LuO₂ rock salt layers [1,2]. The resulting iron sublattice arrangement consists of three stacked triangular bilayers with an equal amount of Fe²⁺ and Fe³⁺ in the unit cell, and this makes the system geometrically frustrated in both the charge and spin channels. Even though intensive experimental and theoretical efforts to clarify the underlying pattern of spin and charge order have been made, its exact nature is still controversial [8].

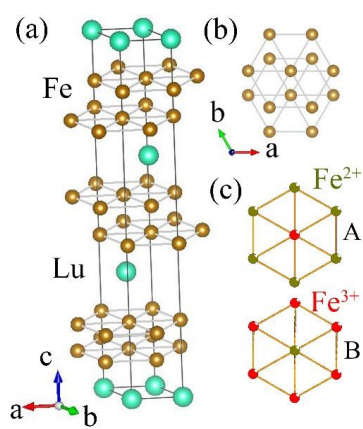

**Figure 1.** (**a**) The crystal structure of LuFe$_2$O$_4$. (**b**) Two adjacent triangular FeO$_5$ layers shifted by $\sqrt{3}/3$ a within the (*ab*) plane. (**c**) The Fe$^{2+}$-(A) and Fe$^{3+}$-rich (B) monolayer charge models. Oxygen ions are omitted for clarity.

A physical picture of the charge order, overcoming the charge frustration in the triangular layers has been proposed in the so called AB model [2,3,5] where A represents an upper iron layer with Fe$^{3+}$ cations surrounded by a honeycomb lattice of Fe$^{2+}$ cations and a lower B layer has the opposite Fe$^{2+}$/Fe$^{3+}$ arrangement (see Figure 1c). Whereas the AB model has been used to account for the ferroelectricity observed in LuFe$_2$O$_4$ due to the net electric dipole moment along the *c*-axis [2,3], X-ray scattering experiments have suggested an antiferroelectric AB–BA bilayer stacking [6,9]. Later, several investigations have found that ferroelectric order can be induced based on this antiferroelectric model when an external electric field is applied [10–12]. A charge order model with AA–BB stacking has also been recently demonstrated through X-ray diffraction study, inconsistent with charge-ordering-based ferroelectricity [7].

On the other hand, the exact spin configuration is also under debate. The magnetic properties of LuFe$_2$O$_4$ are governed by the Ising character of the iron spin in the triangular lattice essentially due to spin-orbit coupling [8,13]. Both two-dimensional (2D) [1] and 3D magnetic structures [12,14] have been previously found in different samples, reflecting the important role played by oxygen stoichiometry [15–18]. In addition, two magnetic phase transitions have been observed by Christianson et al. via single crystal neutron diffraction study. The first transition, observed below T$_N$ = 240 K, is characterized by finite magnetic correlations within ferrimagnetic (FIM) monolayers stacked ferromagnetically along the c direction; while at the second transition (T$_L$ = 175 K) a significant broadening of specific magnetic reflections is observed, indicative of an additional decrease in the magnetic correlation length [14,19]. Moreover, a metamagnetic state in the vicinity of T$_N$, consisting of nearly degenerate FIM and antiferromagnetic (AFM) spin order has been proposed [20]. A distinct FIM spin configuration on the bilayers based on the AB charge order model, where all Fe$^{2+}$ spins in the adjacent monolayers are FM, while Fe$^{3+}$ spins are antiparallel to Fe$^{2+}$ rich layers and nearest neighbour Fe$^{3+}$ spins in the Fe$^{3+}$ rich layer are antiparallel, has been put forward through X-ray magnetic circular dichroism [10,12,21], and has been supported by theoretical studies using both density functional theory and Monte Carlo simulations as well as inelastic scattering experiments [21–23]. Apart from this 2D spin arrangement in the bilayer, Mulders et al. have suggested a long range AFM order along the c direction that facilitates the long range order of electric dipole moment, suggesting the presence of spin-charge coupling [12]. A magnetic field vs. temperature phase diagram above 180 K has been recently proposed, which reveals that magnetic field promotes the FIM order from the degenerate AFM–FIM phases [20].

In this work, we present a neutron diffraction experiment on single crystal LuFe₂O₄₊δ under an applied magnetic field along the Ising direction in the temperature range 5–275 K. The striking effect of the magnetic field on the magnetic structure from our experiment together with previous works brings new insights into the complex magnetic field versus temperature phase diagram and reveals clear evidence for a field-induced coupling between the spin and charge orderings.

## 2. Results

Figure 2a shows the magnetic susceptibility of LuFe₂O₄₊δ under zero field cooled (ZFC) and field cooled (FC) conditions measured by applying a magnetic field of 1 T parallel to the *c*-axis. In accordance with previous results, the ferrimagnetic order appears below $T_N = 240$ K [1]. A significant irreversibility between the ZFC and FC data can be seen at around 185 K. However, the second magnetic phase transition at 175 K reported in Reference [14] is absent in this sample. The magnetization measurements are consistent with a sample containing a slight oxygen excess [1,3]. As shown in the inset of Figure 2a, the magnetic hysteresis loop at 200 K is indicative of a net spontaneous moment.

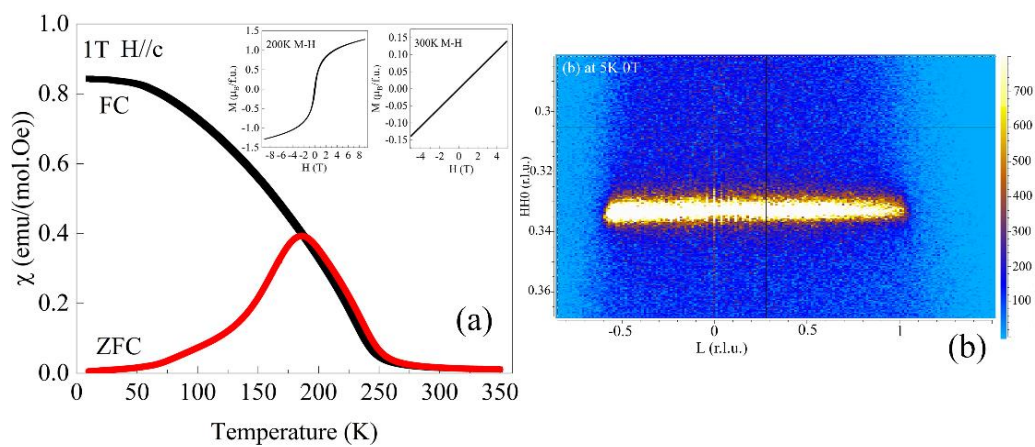

**Figure 2.** (**a**) Magnetic susceptibility of LuFe₂O₄₊δ measured with an applied field of 1 T parallel to the *c*-axis. Insets show the magnetic hysteresis loops at 200 and 300 K. (**b**) The strong spin correlations along 1/3 1/3 L rod at 5 K and 0 T. The absence of intensity when L < −0.6 and >1 is an artefact of the magnet coverage.

The neutron diffraction data reveals strong magnetic diffuse scattering, as can be seen in Figure 2b. This scattering is sharp in the HK0 plane, located in the 1/3 1/3 position, and diffuse along the L direction forming a rod in the reciprocal space. This scattering persists down to 5 K in the zero field and is indicative of 2D magnetic correlations in the FeO₅ triangular bilayers. As shown in Figure 3a, the integrated intensity of the reciprocal space cut along the 1/3 1/3 L direction becomes more intense below 240 K, in good agreement with the magnetization measurement. Remarkably, a long range spin order associated with a propagation vector $k_1 = 0$ and coupled to the 2D spin correlations is unveiled below 240 K, and is characterized by sharp magnetic contribution to the 110, 011 and −221 nuclear reflections, as shown in Figure 3b. This $k_1 = 0$ component is related to a ferromagnetic contribution with the *R-3m′* symmetry as can be appreciated in Figure A1 (an AFM arrangement with the *R-3′m′* symmetry would result in null magnetic intensity in the HK0 plane reflections). A net moment for a single triangular unit of Ising spins is naturally associated with up–up–down (down–down–up) spin configurations. The presence of the coherent scattering with $k_1 = 0$ implies, in the simplest case, an averaging between up–up–down, up–down–up and down–up–up configurations for each triangle. Practically, it means that the numbers of spins with up and down polarization are not equal in each triangular layer, resulting in a net ferrimagnetic moment. An illustrative example of such a ferrimagnetic layer is shown in Figure 3c. The layers are then stacked along the *c*-axis with a random shift within the *(ab)* plane. The strong 2D magnetic scattering rod together with the 3D ferromagnetic component lead to a physical picture of spin ordering in LuFe$_2$O$_{4+\delta}$ from 5 K to 270 K in which the lack of spin correlations along the c direction coexists with the long range Ising ferrimagnetic order in each triangular monolayer.

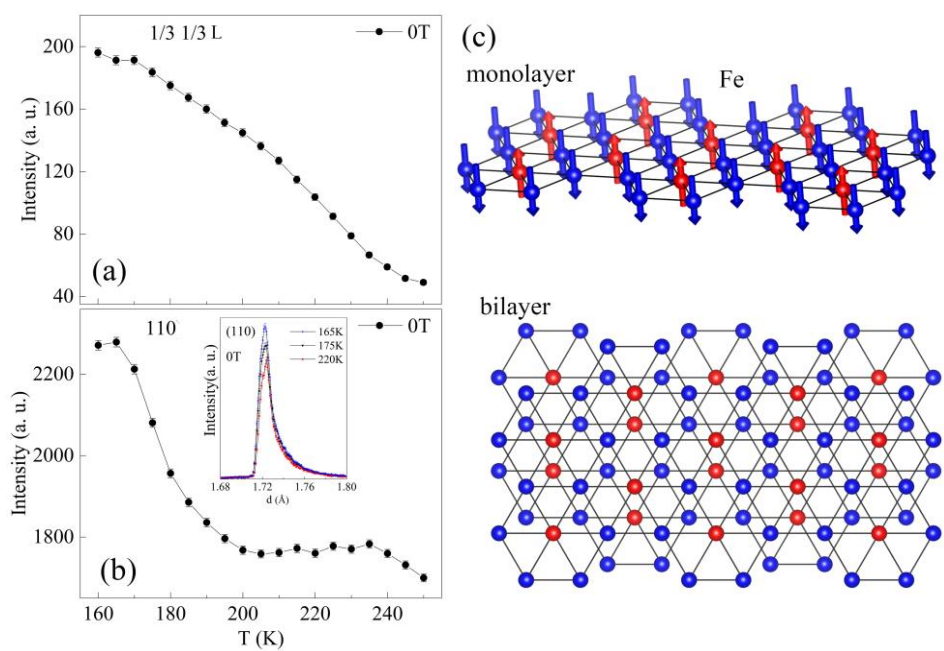

**Figure 3.** Temperature dependence of the integrated intensities of (**a**) the diffuse scattering at 1/3 1/3 L and (**b**) the 110 reflection at 150 K. The inset shows the 110 reflection at 165, 175 and 220 K. (**c**) Schematic drawing of a spin configuration with a net ferrimagnetic moment within a layer (upper panel) and a stacking of two adjacent monolayers into a bilayer (lower panel). Red (resp. blue) spheres represent up (resp. down) spins for the lower diagram.

The magnetic field dependence of neutron magnetic diffraction at 150 K is shown in Figure 4a–c. The application of an external field along the c direction induces a strong change in the distribution of the diffuse scattering. Firstly, there is a strong intensity increase around the 1/3 1/3 1/2 position (Figure 4c), the scattering is peaked at this position but it is still not resolution limited, indicating a finite correlation length. Moreover, a weaker peak-like feature is observed around the 1/3 1/3 0 satellite reflection and interestingly it is no longer centred at the commensurate position in the HK0 plane but is slightly off (as can be appreciated from the inset of Figure 4a,b), indicating the development of incommensurate interactions within the Fe layers. A similar situation is observed in the diffuse scattering at the 2/3 2/3 L position. The observation of Bragg reflections at the 1/3 1/3 1/2 and 2/3 2/3 1/2 as well as other non-integer L positions clearly indicates that the application of an external field promotes 3D spin correlations between the FeO$_5$ triangular bilayers with a propagation vector $\mathbf{k_2}$ = (1/3, 1/3, 3/2) with respect to the parent *R-3m* unit cell. It should be pointed out that this propagation vector is identical to the structural modulation associated with the charge ordering reported in Reference [6] and confirmed in the present study, as can be appreciated by the observation of charge satellite reflections in the high Q region in Figure A2. As presented in Figure 4d,e, the temperature dependence of integrated intensities of 1/3 1/3 0 and 1/3 1/3 1/2 satellite reflections under 5 T shows the onset of spin ordering at 260 K, slightly higher than that under 0 T.

Another outstanding consequence of the magnetic field is the dramatic decrease of integrated intensities of 110, 011 and −221 reflections. As seen in the inset of Figure 4c, the intensity of 110 peak monotonically decreases with increasing magnetic field. Under a magnetic field of 5 T, as shown in Figure 4f, there is no obvious variation of the integrated intensity of the 110 reflection with temperature. These observations show that the external magnetic field suppresses the $\mathbf{k_1}$-related component. It is also important to stress that the magnetic scattering corresponding to the $\mathbf{k_2}$ propagation vector cannot simply be ascribed to a redistribution of the diffuse scattering intensity but instead includes an extra component, likely coming from the magnetic scattering related to the $k_1$ propagation vector in zero field.

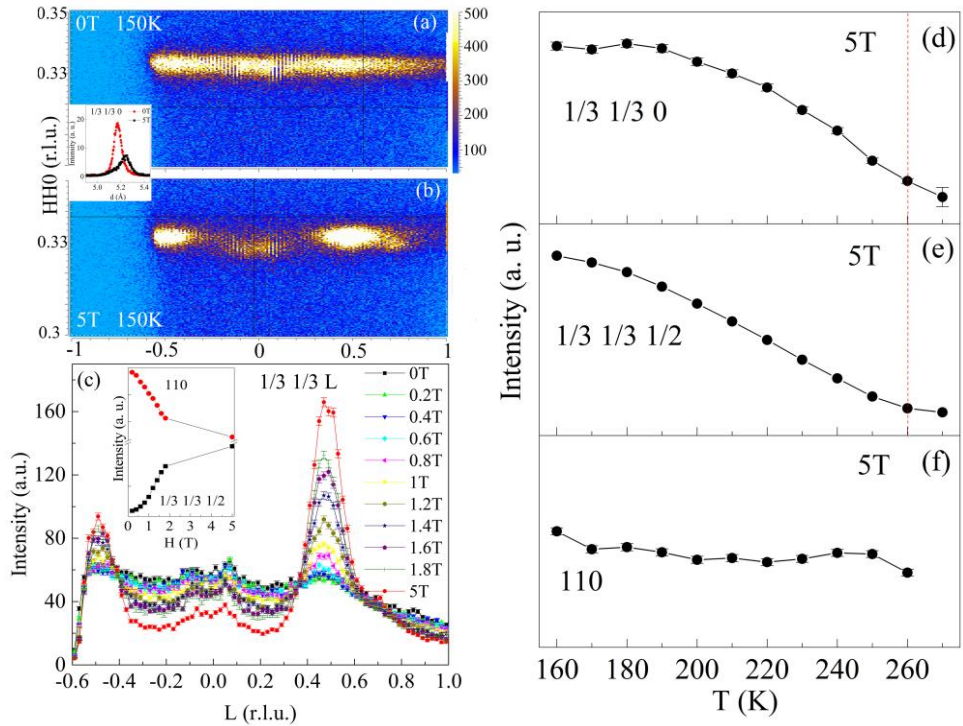

**Figure 4.** (**a**,**b**) Neutron scattering rod along 1/3 1/3 L direction under 0 and 5 T at 150 K. The inset shows the incommensurate character of 1/3 1/3 0 reflection when magnetic field of 5 T is applied along the c-axis. (**c**) Magnetic field dependence of neutron scattering along 1/3 1/3 L direction at 150 K. The inset demonstrates the magnetic field dependence of the integrated intensities around the 1/3 1/3 1/2 position and the 110 reflections at 150 K. (**d**–**f**) Temperature dependencies of the integrated intensities of the 1/3 1/3 0, 1/3 1/3 1/2 and 110 reflections at 150 K under 5 T.

## 3. Discussion

The experimentally observed behaviour can be understood as a magnetic-field-imposed coupling between the spin and charge degrees of freedom. The component of the magnetic field along the c-axis transforms as the time-odd, one-dimensional representation $\Gamma_{2^+}(H_-)$ of the parent *R-3m* space group. The symmetry of the charge order cannot be uniquely identified from the available diffraction data and the only solid experimental evidence is that the propagation vector for this type of distortion is (1/3, 1/3, 3/2), in agreement with other experimental results [6]. There are two six-dimensional irreducible representations associated with this propagation vector whose matrix operators for the generators of the *R-3m* space group are specified in Table 1. Whichever of them is responsible for the symmetry breaking related to the charge ordering, there is always a way to form a tri-linear free-energy coupling term with a time-odd quantity (magnetic order) with the same periodicity as the charge ordering. Specifying the relevant order parameters as $(\eta_+, \eta^*_+, \xi_+, \xi^*_+, \rho_+, \rho^*_+)$ for the charge ordering and $(\eta_-, \eta^*_-, \xi_-, \xi^*_-, \rho_-, \rho^*_-)$ for the field-induced magnetic order, the free-energy invariant reads:

$$H_-(\eta_-\xi^*_+ + \xi_-\rho^*_+ + \rho_-\eta^*_+ + \eta_+\xi^*_- + \xi_+\rho^*_- + \rho_+\eta^*_-) \qquad (1)$$

The subscripts and asterisk indicate time-parity and complex conjugation, respectively. If the charge ordering transforms as the time-even Y1 representation, then the field-induced magnetic ordering must belong to the time-odd Y2 representation. On the contrary, if the time-even Y2 is the relevant representation for the charge ordering, then the time-odd Y1 describes the transformational properties of the induced magnetic order. Thus the condition to couple the charge and spin degrees of freedom in the applied magnetic field is that these order parameters must be transformed by the different representations. At a microscopic level, this implies that the combination of charge and spin orderings with the symmetries of the Y1 and Y2 representations should result in magnetic structures with uncompensated (ferrimagnetic) moments which can be coupled to the external magnetic field. Another conclusion arising from the symmetry consideration is that the coupling invariant does not vanish only if at least two (out of the three) arms of the (1/3, 1/3, 3/2)-propagation vector star are involved. Actually the term is maximal when the full star participates in the symmetry breaking and, therefore, if the charge ordering taking place in the zero-field exploits only a single arm (as for instance it was suggested in Reference [7]), then the application of the magnetic field should result in a re-distribution of the charge, making the ordering multi-*k*. Unfortunately, the current geometry of the diffraction experiment and the restrictions coming from the magnet do not allow us to verify this conclusion.

**Table 1.** Irreducible representations for generating symmetry elements of the *R-3m* space group, associated with the propagation vector $\mathbf{k_2}$ = (1/3, 1/3, 3/2).

| Irrep | {3+│0,0,0} | {2ₓₓ│0,0,0} | {−1│0,0,0} |
|---|---|---|---|
| $\Gamma_{2^+}$ | 1 | −1 | 1 |
| Y1 | $\begin{pmatrix} 0&0&1&0&0&0 \\ 1&0&0&0&0&0 \\ 0&1&0&0&0&0 \\ 0&0&0&0&0&1 \\ 0&0&0&1&0&0 \\ 0&0&0&0&1&0 \end{pmatrix}$ | $\begin{pmatrix} 1&0&0&0&0&0 \\ 0&0&1&0&0&0 \\ 0&1&0&0&0&0 \\ 0&0&0&1&0&0 \\ 0&0&0&0&0&1 \\ 0&0&0&0&1&0 \end{pmatrix}$ | $\begin{pmatrix} 0&0&0&1&0&0 \\ 0&0&0&0&1&0 \\ 0&0&0&0&0&1 \\ 1&0&0&0&0&0 \\ 0&1&0&0&0&0 \\ 0&0&1&0&0&0 \end{pmatrix}$ |
| Y2 | $\begin{pmatrix} 0&0&1&0&0&0 \\ 1&0&0&0&0&0 \\ 0&1&0&0&0&0 \\ 0&0&0&0&0&1 \\ 0&0&0&1&0&0 \\ 0&0&0&0&1&0 \end{pmatrix}$ | $\begin{pmatrix} -1&0&0&0&0&0 \\ 0&0&-1&0&0&0 \\ 0&-1&0&0&0&0 \\ 0&0&0&-1&0&0 \\ 0&0&0&0&0&-1 \\ 0&0&0&0&-1&0 \end{pmatrix}$ | $\begin{pmatrix} 0&0&0&1&0&0 \\ 0&0&0&0&1&0 \\ 0&0&0&0&0&1 \\ 1&0&0&0&0&0 \\ 0&1&0&0&0&0 \\ 0&0&1&0&0&0 \end{pmatrix}$ |

| Irrep | {1│1,0,0} | {1│0,0,1} |
|---|---|---|
| $\Gamma_{2^+}$ | 1 | 1 |
| Y1 | $\begin{pmatrix} e^{-\frac{2}{3}\pi i}&0&0&0&0&0 \\ 0&e^{\frac{2}{3}\pi i}&0&0&0&0 \\ 0&0&e^{-\frac{2}{3}\pi i}&0&0&0 \\ 0&0&0&e^{\frac{2}{3}\pi i}&0&0 \\ 0&0&0&0&e^{-\frac{2}{3}\pi i}&0 \\ 0&0&0&0&0&e^{\frac{2}{3}\pi i} \end{pmatrix}$ | $\begin{pmatrix} -1&0&0&0&0&0 \\ 0&-1&0&0&0&0 \\ 0&0&-1&0&0&0 \\ 0&0&0&-1&0&0 \\ 0&0&0&0&-1&0 \\ 0&0&0&0&0&-1 \end{pmatrix}$ |
| Y2 | $\begin{pmatrix} e^{-\frac{2}{3}\pi i}&0&0&0&0&0 \\ 0&e^{\frac{2}{3}\pi i}&0&0&0&0 \\ 0&0&e^{-\frac{2}{3}\pi i}&0&0&0 \\ 0&0&0&e^{\frac{2}{3}\pi i}&0&0 \\ 0&0&0&0&e^{-\frac{2}{3}\pi i}&0 \\ 0&0&0&0&0&e^{\frac{2}{3}\pi i} \end{pmatrix}$ | $\begin{pmatrix} -1&0&0&0&0&0 \\ 0&-1&0&0&0&0 \\ 0&0&-1&0&0&0 \\ 0&0&0&-1&0&0 \\ 0&0&0&0&-1&0 \\ 0&0&0&0&0&-1 \end{pmatrix}$ |

Finally, it is worth noting that following the discussion presented above, the presence of magnetic scattering in the incommensurate position close to 1/3 1/3 0 might indicate the existence of a minor fraction of incommensurate charge ordered phase. We however do not have any experimental evidence for a structural modulation associated with this periodicity in our sample.

## 4. Experiments and Methods

The sample used here was cleaved from high-quality single crystal grown by the floating zone method, and the detailed experimental procedure can be found in Reference [22].

Magnetic susceptibility was measured using a vibrating sample magnetometer (VSM) (Physical Property Measurement System, Quantum Design, San Diego, CA 92121, USA) between 10 K and 350

K under magnetic field of 1 T under both zero-field-cooled (ZFC) and field-cooled (FC) conditions. Magnetic hysteresis loops were measured at 200 and 300 K.

Magnetic-field dependent single crystal neutron diffraction experiments were carried out at the ISIS pulsed neutron and muon facility of the Rutherford Appleton Laboratory (UK), on the WISH diffractometer [24] located at the second target station. The single crystal (~0.15 g) was mounted in a vertical field superconducting cryomagnet with magnetic field up to 13.6 T and measured over the temperature range of 5–270 K with magnetic field applied along the *c*-axis. Group theoretical calculations were done using ISODISTORT [25] and Bilbao Crystallographic Server (Magnetic Symmetry and Applications) software [26]. Simulations of the neutron single crystal data were performed with the help of the Jana2006 software [27].

## 5. Conclusions

We investigated the effect of magnetic field on spin correlations in the charge ordered single crystal of LuFe$_2$O$_{4+\delta}$ through magnetization measurement and neutron diffraction. In the absence of a magnetic field, the spin and charge subsystems are effectively decoupled. The spin forms 2D correlations featured by the magnetic diffuse scattering rod in the 1/3 1/3 L direction and a 3D ferromagnetic component below T$_N$ = 240 K. An external magnetic field suppresses the diffuse scattering and promotes 3D spin correlations with the propagation vector **k**$_2$ = (1/3, 1/3, 3/2). This periodicity is common for both the magnetic and charge-ordered sublattices and a combination of these experimental results with symmetry consideration provides evidence of a magnetic field imposed coupling between the spin and charge degrees of freedom.

**Acknowledgments:** Lei Ding thanks support from the Rutherford International Fellowship Programme (RIFP). This project has received funding from the European Union's Horizon 2020 research and innovation programme under the Marie Skłodowska-Curie Grant Agreements No. 665593 awarded to the Science and Technology Facilities Council. The work at the University of Warwick was supported by EPSRC, UK through Grant M028771/1. We thank the support received during material characterizations in the Materials characterization laboratory at the ISIS facility.

**Author Contributions:** Pascal Manuel, Dmitry D. Khalyavin and Lei Ding conceived the work; Andrew T. Boothroyd, Dharmalingam Prabhakaran and Geetha Balakrishnan performed single crystal growth; Pascal Manuel and Dmitry D. Khalyavin performed the neutron diffraction experiments; Lei Ding, Fabio Orlandi, Dmitry D. Khalyavin and Pascal Manuel analyzed the neutron data; Lei Ding also contributed to the magnetization measurements; Lei Ding, Fabio Orlandi, Dmitry D. Khalyavin and Pascal Manuel wrote the paper.

**Conflicts of Interest:** The authors declare no conflict of interest.

## Appendix A

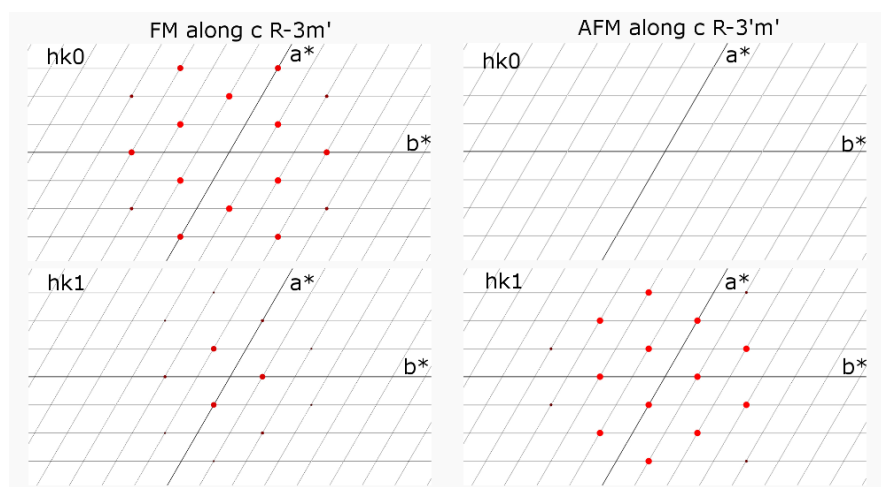

**Figure A1.** Simulated magnetic reflections in HK0 (**Top**) and HKL (**Bottom**) plane for a ferromagnetic (**Left**) and antiferromagnetic order (**Right**) along the c direction, the radius and the brightness of the reflections are an indication of the magnetic structure factor. The simulations are performed with the Jana2006 software.

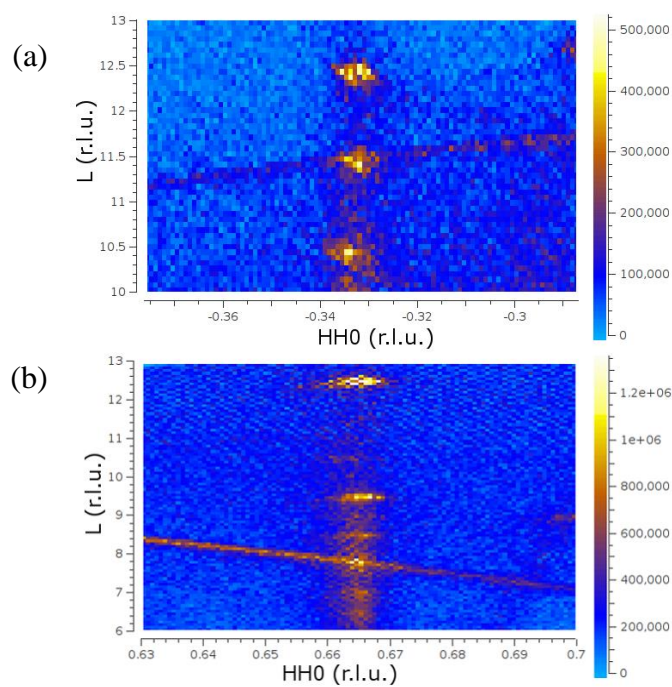

**Figure A2.** Neutron scattering in the high Q region at the (a) 1/3 1/3 L and (b) 2/3 2/3 L positions showing the presence of charge ordering satellite reflections ascribable to the **q** = (1/3, 1/3, 3/2) modulation vector.